\documentclass[aps,prd,preprint,showpacs,superscriptaddress,amsmath]{revtex4}

\usepackage{longtable}
\usepackage{graphicx}
\usepackage{amsmath,amssymb}
\usepackage{color}
\usepackage{bbm}

\newcommand{\beq}{\begin{equation}}
\newcommand{\eeq}{\end{equation}}
\newcommand{\bdm}{\begin{displaymath}}
\newcommand{\edm}{\end{displaymath}}

\definecolor{Gray}{gray}{0.9}

\graphicspath{{./plots/}}
\begin{document}

\title{Global characterization of seismic noise with broadband seismometers}
\author{M Coughlin$^1$ and J Harms$^2$}
\affiliation{$^1$Carleton College, Northfield, MN 55057, USA}
\affiliation{$^2$California Institute of Technology, Pasadena, California 91125, USA}

\begin{abstract}
In this paper, we present an analysis of seismic spectra that were calculated from all broadband channels (BH?) made available through IRIS, NIED F-net and Orfeus servers covering the past five years and beyond. A general characterization of the data is given in terms of spectral histograms and data-availability plots. We show that the spectral information can easily be categorized in time and regions. Spectral histograms indicate that seismic stations exist in Africa, Australia and Antarctica that measure spectra significantly below the global low-noise models above 1\,Hz. We investigate world-wide coherence between the seismic spectra and other data sets like proximity to cities, station elevation, earthquake frequency, and wind speeds. Elevation of seismic stations in the US is strongly anti-correlated with seismic noise near 0.2\,Hz and again above 1.5\,Hz. Urban settlements are shown to produce excess noise above 1\,Hz, but correlation curves look very different depending on the region. It is shown that wind speeds can be strongly correlated with seismic noise above 0.1\,Hz, whereas earthquakes produce seismic noise that shows most clearly in correlation around 80\,mHz. 
\end{abstract}

\maketitle

\section{Introduction}
\label{sec:Intro}

Seismic spectra in the range 10\,mHz to 10\,Hz are influenced by a variety of sources including anthropogenic and atmospheric disturbances, earthquakes and ocean waves \cite{BCB2006}. Whereas spectral studies can be easily developed for smaller seismic networks and sufficiently short periods of time, a systematic attempt to link spectra and sources on a global scale using many years of data is challenging. One example of a global study of seismic spectra is the calculation of seismic noise models \cite{Pet1993,BDE2004,McEA2009}. Spectral data have also been used to characterize seismicity over larger regions in more detail \cite{McBu2004,BrEA2009}. More recently, results from a study of temporal variations of seismic spectra have been reported in \cite{RiEA2010}. Even though some links between seismicity and sources are well established, like excess noise above 1\,Hz and proximity to populated areas, it is also true that there is no general model that could explain seismicity within certain frequency bands globally. The notable exception is the oceanic microseism around 0.3\,Hz that seems to dominate seismic ground spectra everywhere on Earth \cite{HMS1963,ToLa1968,Ces1994,FKK1998}. 

In this paper, we will extend previous studies by including data from a significantly larger number of stations, and by calculating temporal and spatial correlations between seismic spectra and auxiliary data: population, wind speeds, topography and earthquake events. A global analysis of seismicity is presented based on publicly available broadband data (BH? channels) from IRIS, NIED F-net, and Orfeus servers. The time-resolution of the spectral data used in this paper is 3 hours, which also allows us to investigate diurnal variations. For all years included in this study, density of seismic stations was high in the US, Europe and Japan. Within the last years, a significant part of all available broadband data in the US were recorded by hundreds of stations forming the high-density Transportable Array moving from the west to the east (see for example \cite{BuEA2012}). 

The paper is organized as follows. In section \ref{sec:seismicdata}, we discuss the origin and acquisition of the seismic data used in this paper and give a general characterization in terms of data availability. In section \ref{sec:NoiseModels}, we present global noise models taking into account station density. In section \ref{sec:Correlations}, we correlate the seismic spectra with population, earthquake, and wind-speed data. Our conclusions are summarized in section \ref{sec:Conclude}.

\section{Data Processing}
\label{sec:seismicdata}

We have acquired seismic data that are publicly available through servers in Japan (F-net), Europe (Orfeus) and the US (IRIS). Further details are given in table \ref{tab:SeismicDataTable}. We systematically downloaded and processed data from all stations with channel names BH?. Stations are supplied with Nanometrics T240, Streckeisen STS-1/STS-2, G\"uralp CMG-3T, and Geotech KS-54000 broadband seismometers. Data are available for some of these stations as far back as the early 1970's through the present. The analysis in this paper is based on data recorded between the years 2007 -- 2011, with comparatively minor contributions from years before 2007. In the future, it will be possible to continue the systematic download and processing of data year by year going further into the past allowing us to carry out long-term studies of seismic data.

\begin{table}

  \begin{center}
  \small
  \begin{tabular}{|l|c|c|}
  \hline
\hspace{0.2cm} Network/Server & Number of Stations & Predominant Location \\
  \hline
IRIS (without TA) & 2143 & all continents \\
IRIS, TA & 1198 & USA \\
Orfeus & 289 & mainly Europe \\
F-net & 74 & Japan \\
\hline
  \end{tabular}
  \end{center}
 
 \caption{Table of seismic networks/servers, locations of seismic stations. The web links for download are www.iris.edu/ws/bulkdataselect/query (IRIS), ftp://www.orfeus-eu.org/pub/data/continuous (Orfeus), and http://www.fnet.bosai.go.jp/REGS/dataget (F-net, login required). The Transportable Array (TA) is listed separately since it comprises a large number of non-permanent stations that are being moved since 2007 from the west coast of the US towards the east. In the future, when the systematic download and processing of data before 2007 from all three servers is continued, the number of stations included in our dataset is likely to increase.}
 \label{tab:SeismicDataTable}
\end{table}

\begin{figure}[t]
\hspace*{-0.5cm}
\includegraphics[width=2.9in]{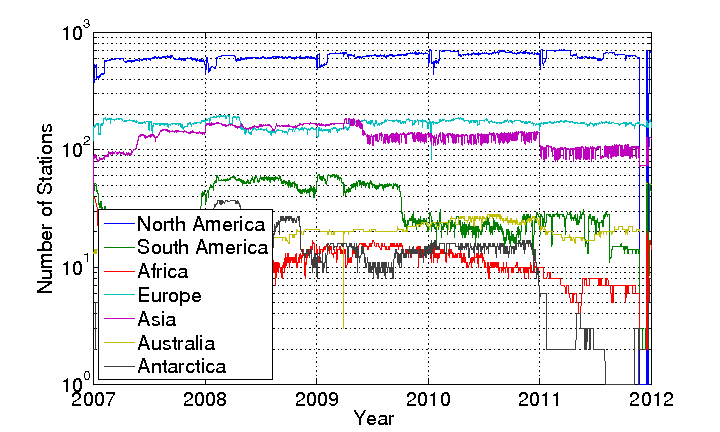}
 \includegraphics[width=2.9in]{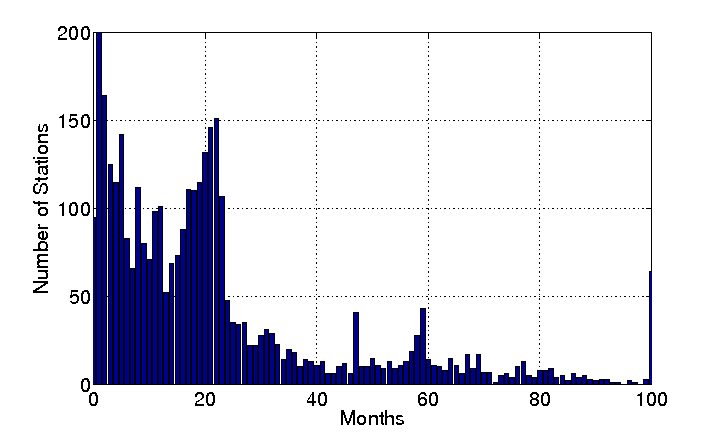}
 \caption{Left: The plot shows the number of available stations for each continent as a function of time since 2007. As the download of data older then 2007 is incomplete, the curves drop rapidly before 2007. Right: The histogram of station numbers show that many stations provide less than a year of data. The peak around 20 months is produced by TA stations, which have a lifetime of about 1.5 years. The peaks at 48 and 60 months are artifacts since the systematic download of data is only complete for the past 5 years.}
 \label{fig:NumOfDays}
\end{figure}

The plots in figure \ref{fig:NumOfDays} give a more detailed view on data availability as a function of time and total amount of data provided by each station. Most stations recorded data for more than one year. The peak in the histogram around 20 months contains the stations of the Transportable Array (TA).

In preparation for this study, seismic spectra were calculated with sufficiently high time resolution so that, for example, diurnal variations or correlations with average wind speeds could be studied more accurately. One data file was generated per station per day for each channel BH? of a 3-axis broadband seismometer. Every data file contains the station name, its latitude/longitude, and the calibration factor of the raw data. The spectra stored in the files are already divided by the calibration factor, which is saved so that we can recalibrate data at a later time if necessary. A day is divided into 8 segments and for each of these 3-hour stretches, we calculated power spectral densities. The spectra are based on 128\,s FFTs using a Nuttall window. Since only 84 of these spectra can be calculated per segment without overlap, percentile curves of the seismic spectra would show high variation. Therefore, we instead decided to calculate two different average spectral densities (with $N=84$):
\beq
S_{\rm lin}=\frac{1}{N}\sum\limits_{i=1}^NS_i, \qquad S_{\rm log}=\exp\left(\frac{1}{N}\sum\limits_{i=1}^N\log(S_i)\right)
\eeq
The logarithmic average provides a spectrum that is less susceptible to occasional strong disturbances. In addition to these average spectra, the mean value of every 3-hour time series is also stored in the file together with a frequency vector and a time vector with starting times for each 3-hour segment. Finally, information is stored about the number of samples that were actually present in each 3-hour stretch.

\begin{figure}[t]
\centerline{
\includegraphics[width=5in]{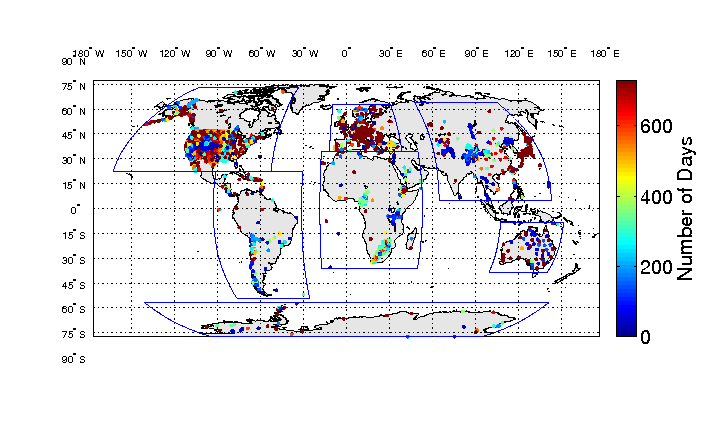}}
\caption{The plot shows the location of seismic stations that provided data analyzed for this paper. The colors indicate the number of days of data obtained from these stations. In the following, when we refer to data from specific continents, then we mean the 7 regions that are indicated in this plot.}
\label{fig:allcontinents}
\end{figure}

The complete distribution of stations that contributed to this study is shown in figure \ref{fig:allcontinents}. The colors indicate the number of days of data recorded by each station.
Station density is exceptionally high in the US, Europe and Japan, but it should be kept in mind that many US stations were only temporarily installed as part of the TA.

\section{Seismic Spectra}
\label{sec:NoiseModels}
Seismic fields are non-stationary and vary significantly from one location to the other. Their properties can depend on distance to populated areas, distance to the coast, local earthquake rate, local average wind speeds, and geological features. We will study some of these links in the next section. In this section, we present spectra averaged over many years and stations. 

\begin{figure}[ht!]
 \begin{center}
Africa \hspace*{2.7cm} Antarctica \hspace*{2.9cm}  Asia\\
 \includegraphics[width=1.7in]{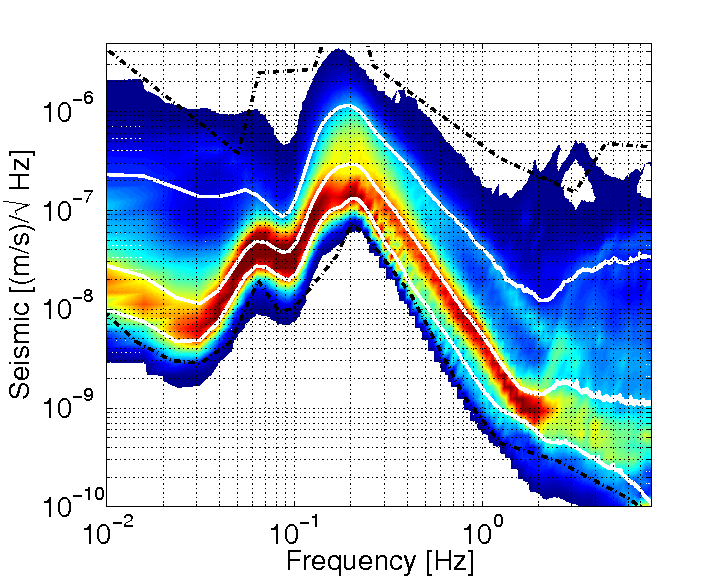}
 \includegraphics[width=1.7in]{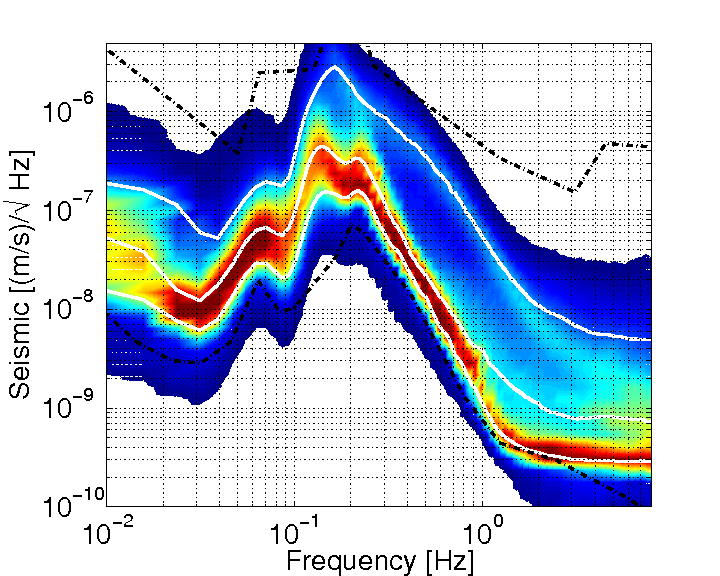}
 \includegraphics[width=1.7in]{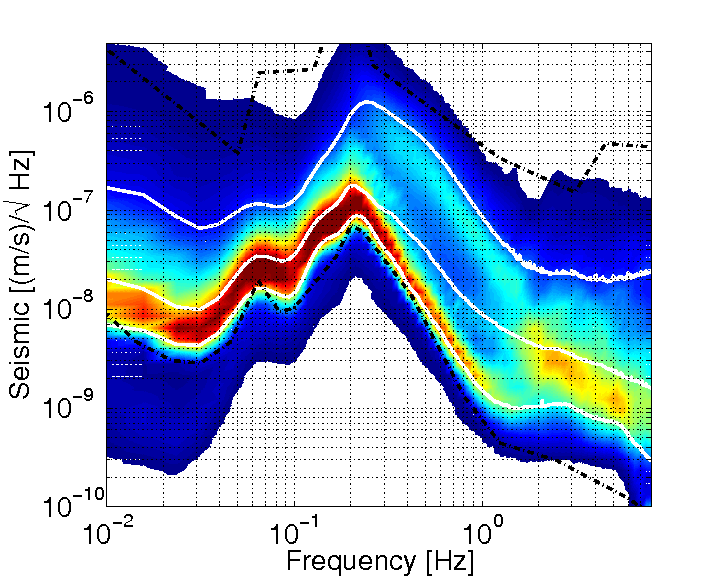}
\hspace*{0.4cm} Australia \hspace*{2.7cm} Europe \hspace*{2.2cm}  North America\\
 \includegraphics[width=1.7in]{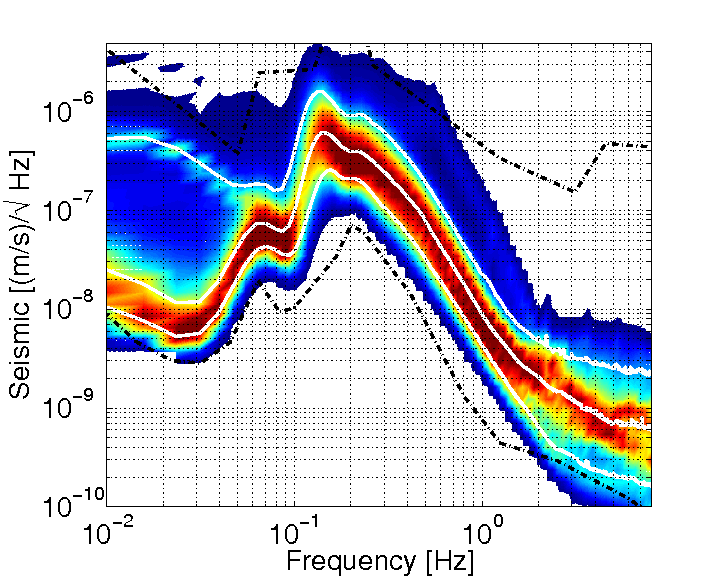}
 \includegraphics[width=1.7in]{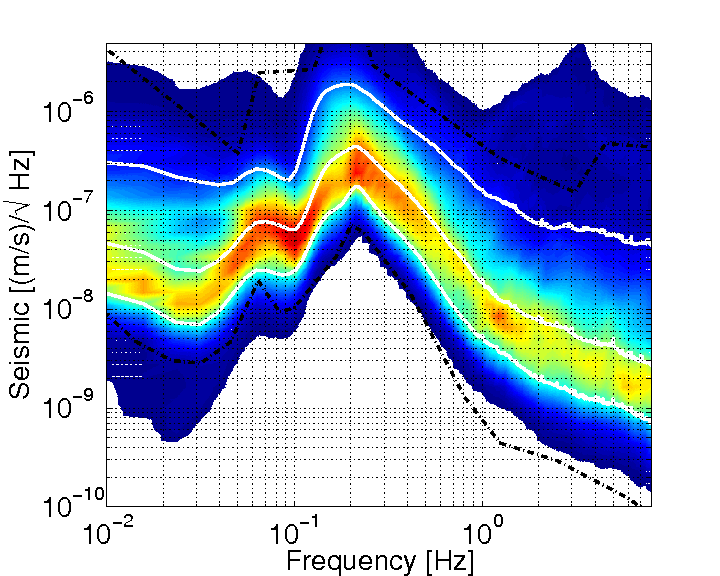}
 \includegraphics[width=1.7in]{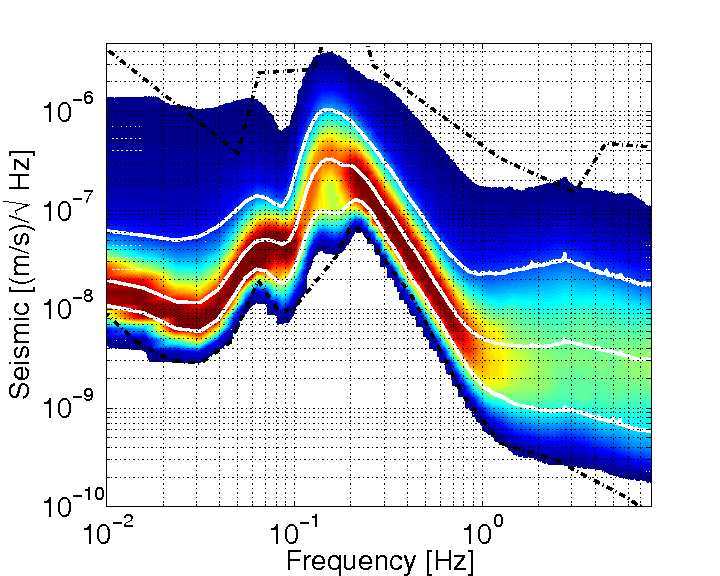}
 South America\\
 \includegraphics[width=1.7in]{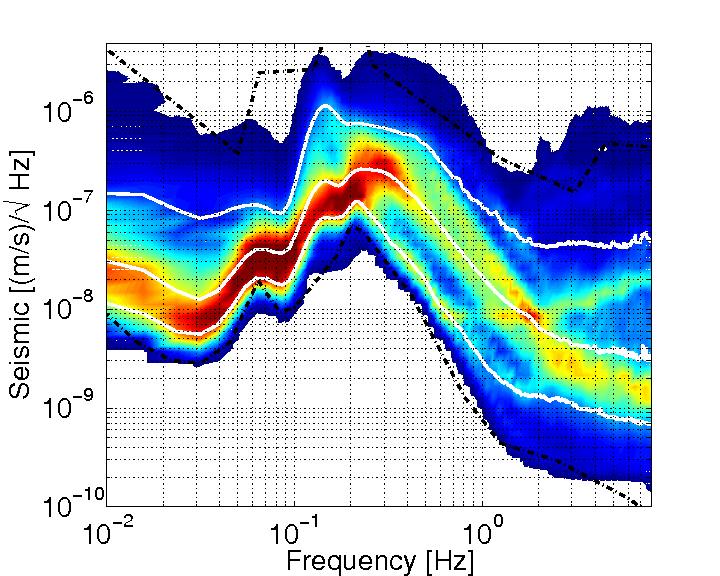}
 \end{center}
 \caption{The plots show histograms of 3-hour spectra of all stations on the respective continent. The dashed, black lines correspond to global low and high-noise models \cite{Pet1993}.  The 10th, 50th, and 90th percentiles computed from the histograms are shown as white solid lines. The number of stations that contributed to these plots are 81 (Africa), 16 (Antarctica), 173 (Asia), 24 (Australia), 148 (Europe), 1369 (North America, many of them temporary being part of TA), and 58 (South America). The histograms contain data from 2007 -- 2012 only. Therefore the station numbers are smaller than in table \ref{tab:SeismicDataTable}.}
 \label{fig:ContinentPDFs}
\end{figure}

We first describe our method to combine spectra from different stations in networks for regional or global plots. In order to account for local station density, we calculate a Delaunay triangulation from station locations for each continent and use triangle areas as weights. The triangulation is carried out after mapping the station coordinates via a Lambert equal-area map projection. The first step is to identify collocated stations (stations closer than 1\,km to each other) to avoid double counting and to obtain better results for the triangulation. Next, the largest 10\% of all triangles in the same region are removed. This step is necessary to avoid artifacts from gaps in networks, especially on continents like Africa or most part of Asia, where a single outlier (e.~g.~a station being close to an urban settlement) could be overemphasized if it is associated with a large triangle area. Each triangle is assigned a spectrum that is the average of spectra of the three stations at its corners. The overall average spectrum is calculated by using the triangle areas as weights. A similar scheme is applied to the spectral histograms that are shown in figure \ref{fig:ContinentPDFs}. Here, the histograms from corner stations of triangles are added to obtain a histogram for the triangle. Then histograms from all triangles are summed again using triangle areas as weights. 
\begin{figure}[t]
\centering
 \includegraphics[width=4in]{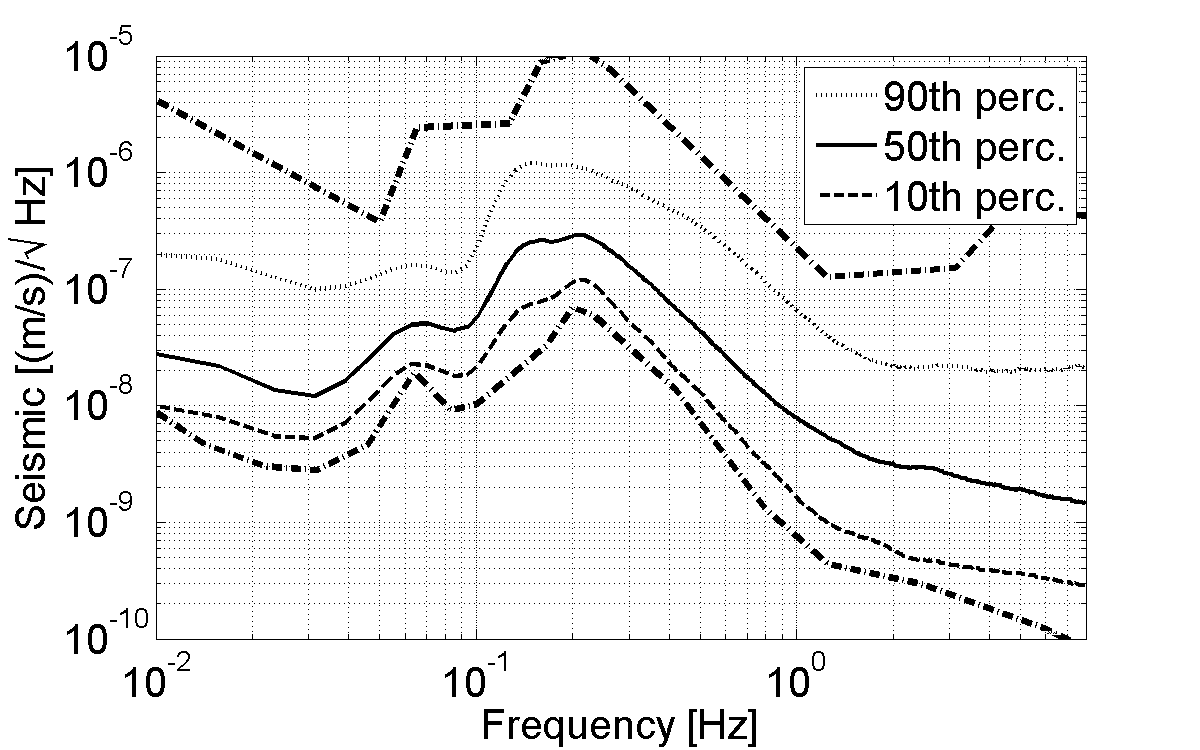}
 \caption{The percentiles are derived from a global spectral histogram containing 5 years of data from all available stations. The 10th percentile lies close to the low-noise model consistent with the results reported in \cite{BDE2004}. The small discrepancy comes from the fact that the percentiles presented here are based on 3-hour averaged spectra. A shorter averaging time would lead to a smaller 10th percentile.}
 \label{fig:NoiseModelComparison}
\end{figure}
The global percentiles shown in figure \ref{fig:NoiseModelComparison} are obtained by combining histograms from all 7 continents using the sum of areas of all triangles on a continent as weighting scheme. The 10th percentile is close to the low-noise model as was already pointed out in \cite{BDE2004}.

\begin{figure}[t]
 \centering
 \includegraphics[width=4in]{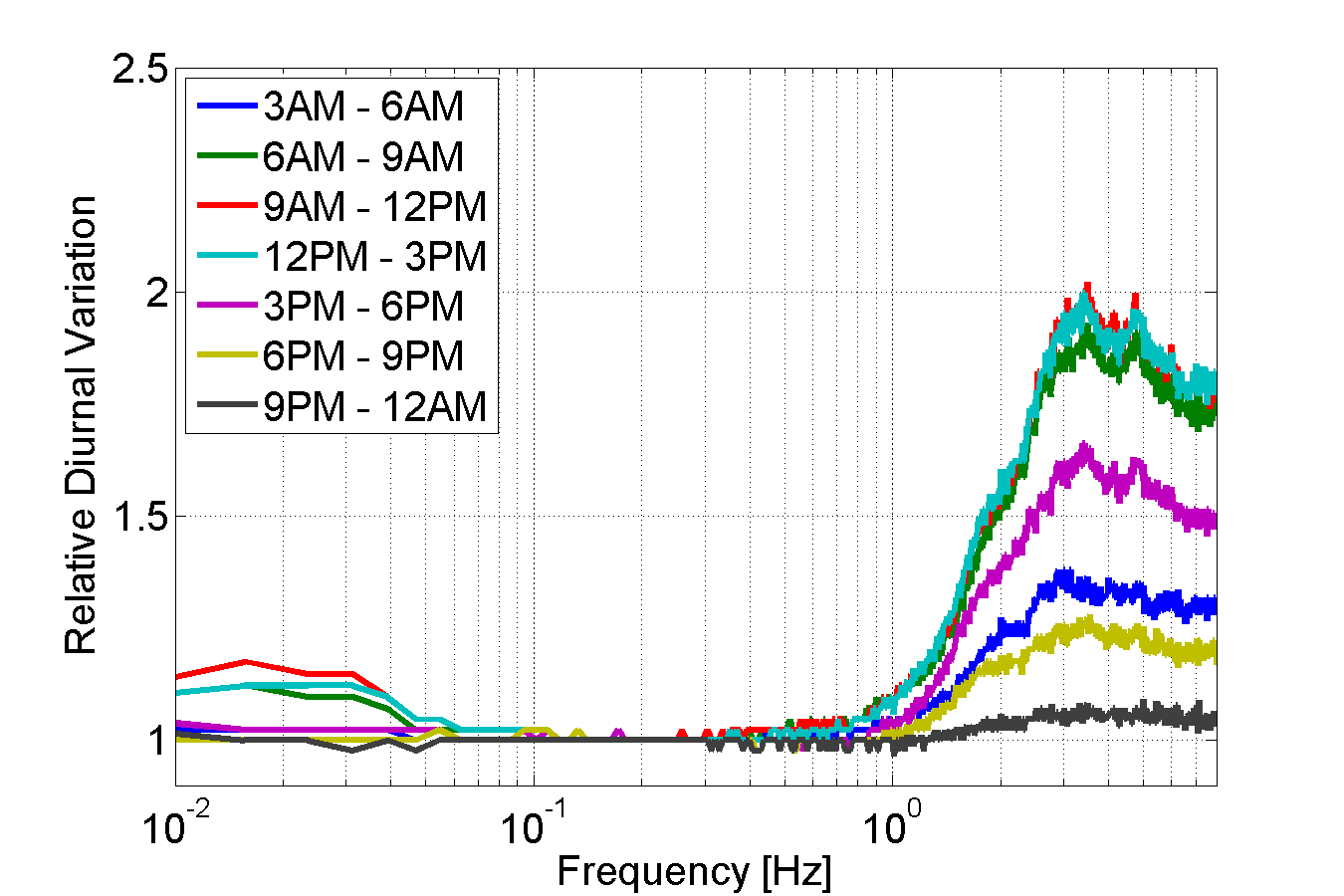}
 \caption{The plot shows global medians within 7 ranges of local station time divided by the midnight median 12am -- 3am. As expected, seismic noise is stronger during the day. Only the frequency range of the oceanic microseisms does not show significant diurnal variation. The absolute values of diurnal variation depend on what type of spectrum one chooses to plot. Variations are about a factor 1.5 stronger in the 90th percentile, and about a factor 1.5 weaker in the 10th percentile.}
 \label{fig:diurnal}
\end{figure}

Due to diurnal variations in factors such as anthropogenic activity or potentially climatic variations, it can be expected that the ambient seismic spectrum be different during the day and at night. To test this, we calculate a local station time according to 
\beq
\mbox{local time} = \mbox{UTC}+\frac{12\,\rm h}{180^\circ}\;\mbox{longitude}
\eeq
and then calculate medians of 3-hour stretches of local time individually over all stations as described previously. The result is shown in figure \ref{fig:diurnal}. All curves are divided by the midnight spectrum 12am -- 3am, which serves as a reference.
As one would expect, seismic noise is larger during the day except for the frequency range of the oceanic microseisms. The absolute value of diurnal variation is different for different types of spectra. For example, the variation becomes larger when going from low to high percentiles. This means that not only the stationary part of the seismic field is stronger, but also that seismic disturbances are more frequent during the day.

\begin{figure}[t]
\hspace*{-0.5cm}
\includegraphics[width=2.9in]{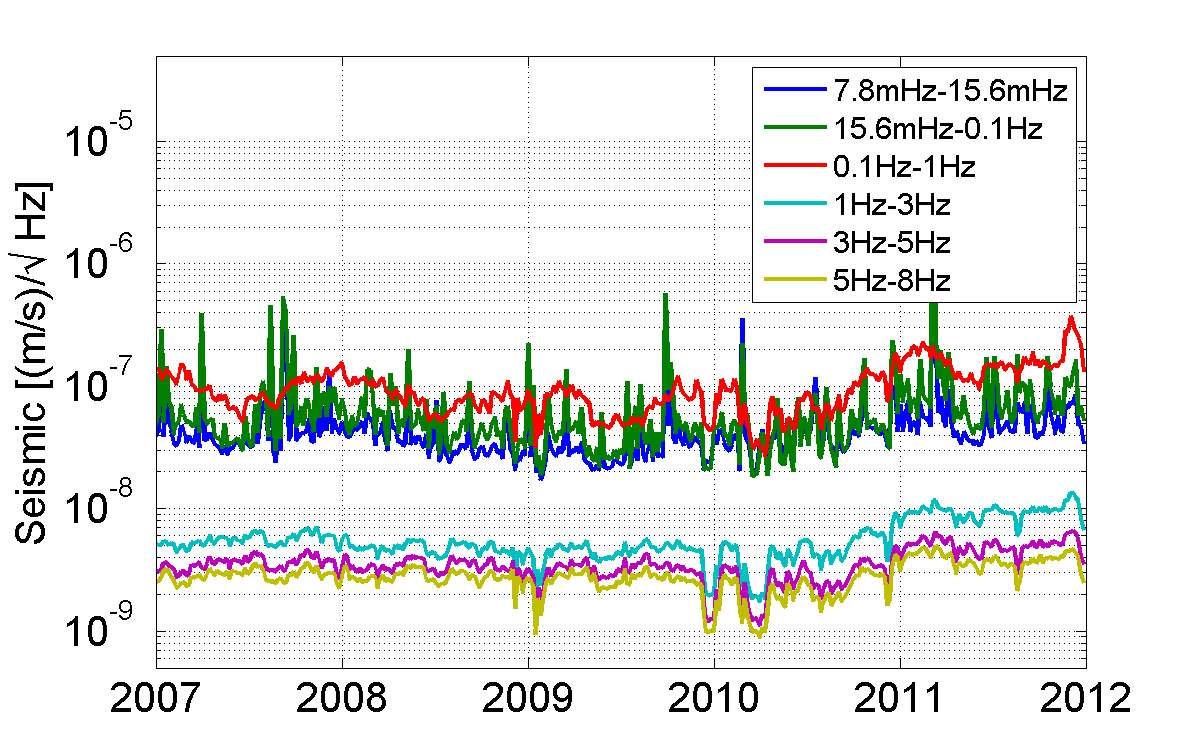}
\includegraphics[width=2.9in]{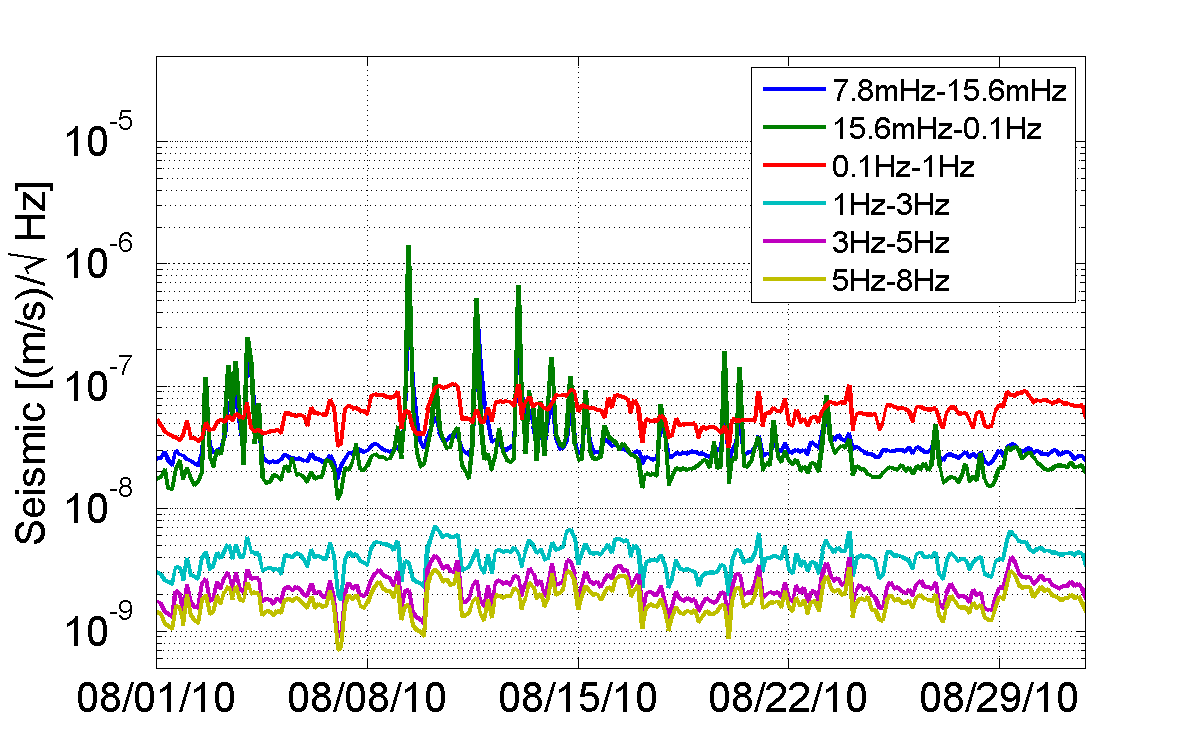}
 \caption{The plot shows the evolution of global seismicity in 6 different frequency bands over the past 5 years. Only stations were included that provided data every month. In this way, a bias is avoided related to the construction or decommissioning of seismic networks. Each data point in the 5-year plot represents the spectral density averaged over 5 days, whereas the August 2010 plot has the full 3-hour resolution. A slight increase in seismicity in all bands since middle of 2010 can be observed.}
 \label{fig:GlobalSeismicAverage}
\end{figure}

We now seek to understand how the seismic spectra change as a function of time. The basic idea is to construct a time series of spectral values. For this purpose we divide the frequency range into 6 intervals. The frequency boundaries were chosen so that the seismic sources in each interval have common characteristics (for the majority of stations). The first interval, 7.8\,mHz - 15.6\,mHz, contains the lowest frequencies of our spectra that typically have no contributions from (primary) oceanic microseisms. These contributions typically lie within the second interval 15.6\,mHz - 0.1\,Hz. The secondary oceanic microseisms dominate seismic noise in the third range 0.1\,Hz - 1\,Hz. The remaining intervals, 1\,Hz - 3\,Hz, 3\,Hz - 5\,Hz, and 5\,Hz - 8\,Hz are chosen more arbitrarily, but a division seems useful since they should each contain seismic waves from sources from varying distance (typically, the higher the frequency, the closer the seismic source). All three intervals are highly influenced by anthropogenic activity or turbulence due to strong winds, which couple to the ground through buildings and trees. The curves are shown in figure \ref{fig:GlobalSeismicAverage} for 2007 -- 2011.
A slight increase in seismicity since middle of 2010 can be observed in all bands. In the future, we desire to study spectral evolution starting further into the past, especially as the 0.1\,Hz -- 1\,Hz band should be strongly linked to climatic trends. For this study, only stations have been included that produced data every month within these 5 years (in other words, data production was continuous, but occasional brief down-times were accepted).

\section{Correlations with seismic spectra}
\label{sec:Correlations}
\begin{figure}[t]
\hspace*{1.9cm}Station elevation\hspace{4.1cm}Urban settlements\\
\hspace*{-0.5cm}
\includegraphics[width=2.9in]{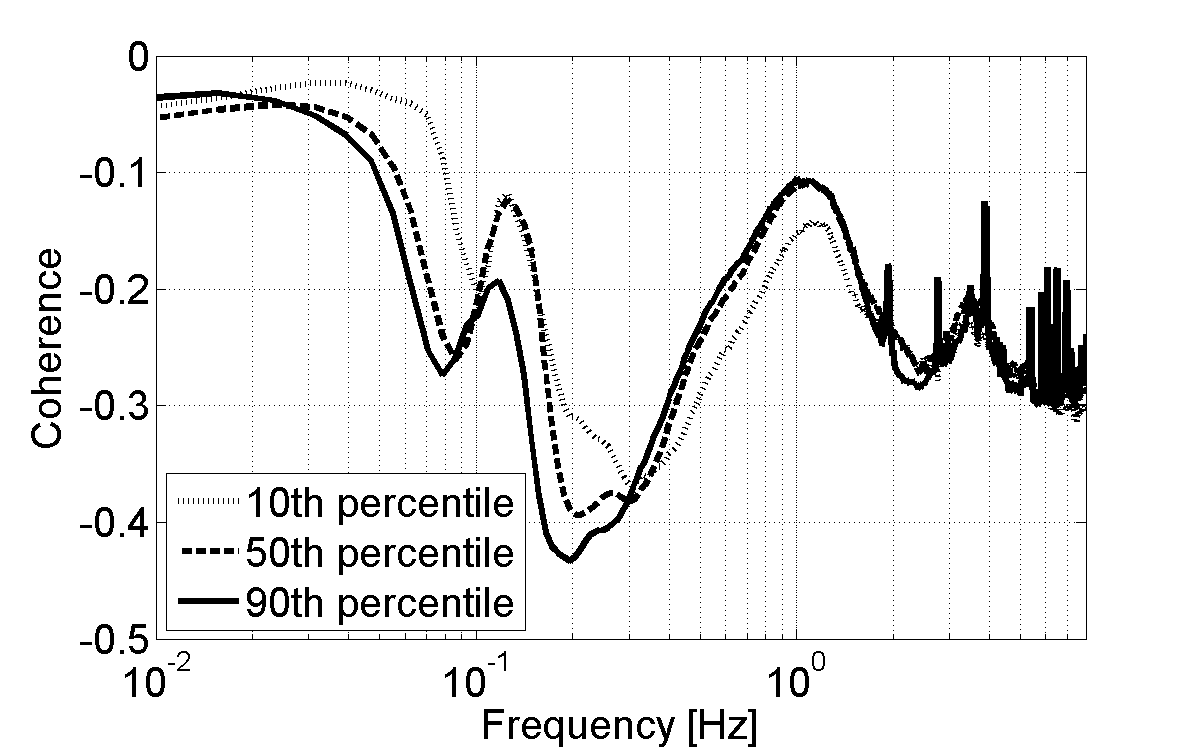}
\includegraphics[width=2.9in]{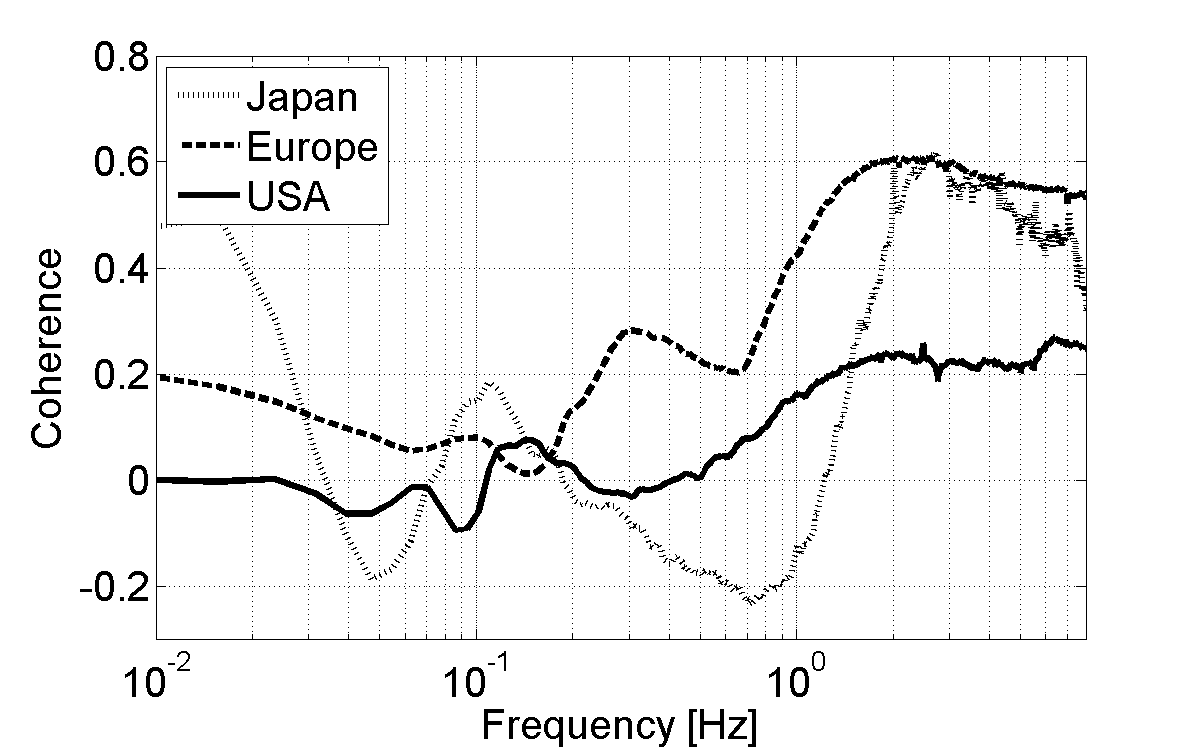}
 \caption{Left: Correlation between station elevation and different percentiles of seismic spectra in the United States vs. frequency. Whereas ambient noise spectra below 50\,mHz do not depend significantly on station elevation, elevation plays an important role for the measured oceanic microseisms and also for the high-frequency regime. Coherence values between 50\,mHz and 1\,Hz are low mostly because high-elevation stations are more distant to the coast. Above 1\,Hz, coherence decreases again since population density is lower in regions of high elevation. Right: Correlation between seismic spectra and population numbers of urban settlements within 50\,km of seismic stations. The coherence is evaluated for the three regions that have high population density and contribute a high number of seismic stations. Above 1\,Hz, seismic fields are strongly influences by anthropogenic activity. As explained in the text, many details of the curves could be related to geographic/geologic features that are themselves correlated with population density, rather than being directly explained by anthropogenic activity.}
 \label{fig:SpatialCorrelation}
\end{figure}

\begin{figure}[t]
\hspace*{2.2cm}Earthquakes\hspace{5cm}Wind speeds\\
\hspace*{-0.5cm}
 \includegraphics[width=2.9in]{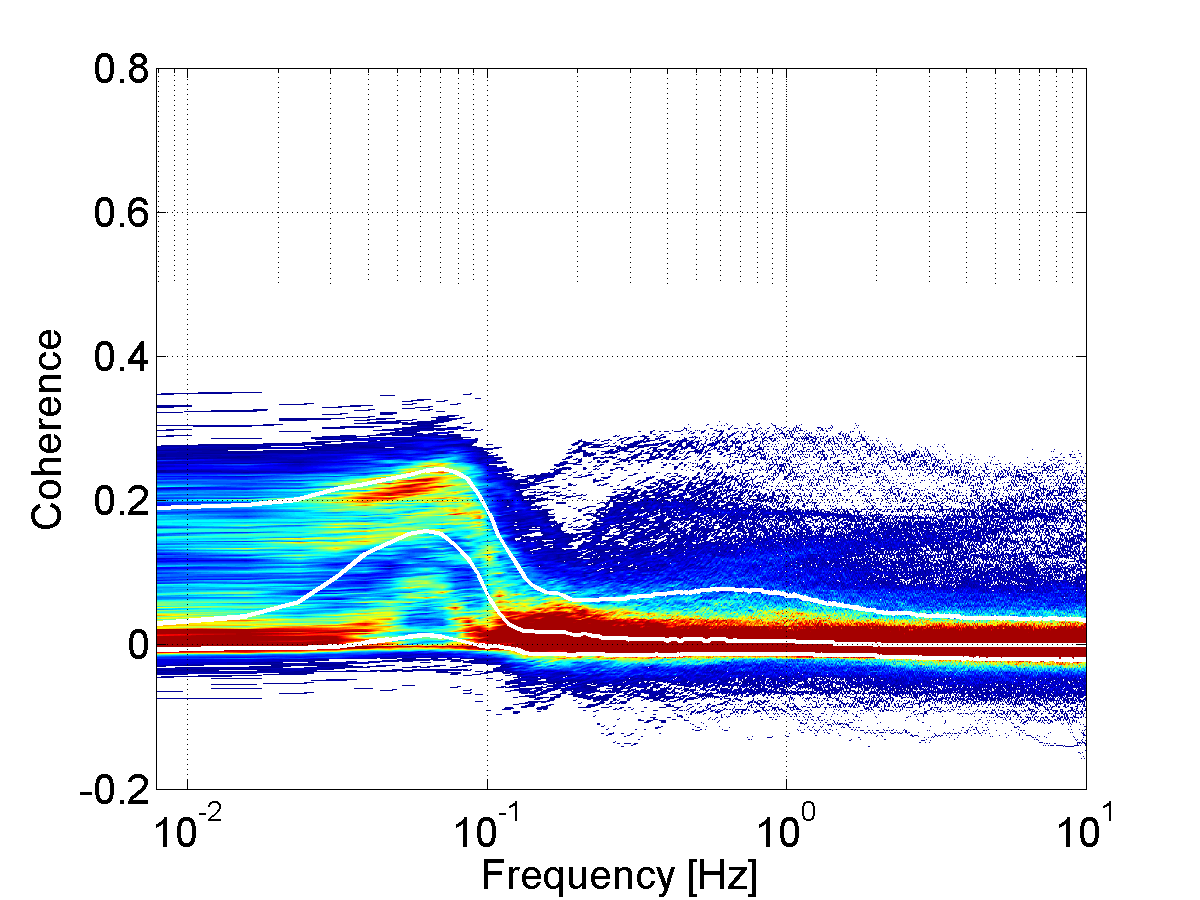}
 \includegraphics[width=2.9in]{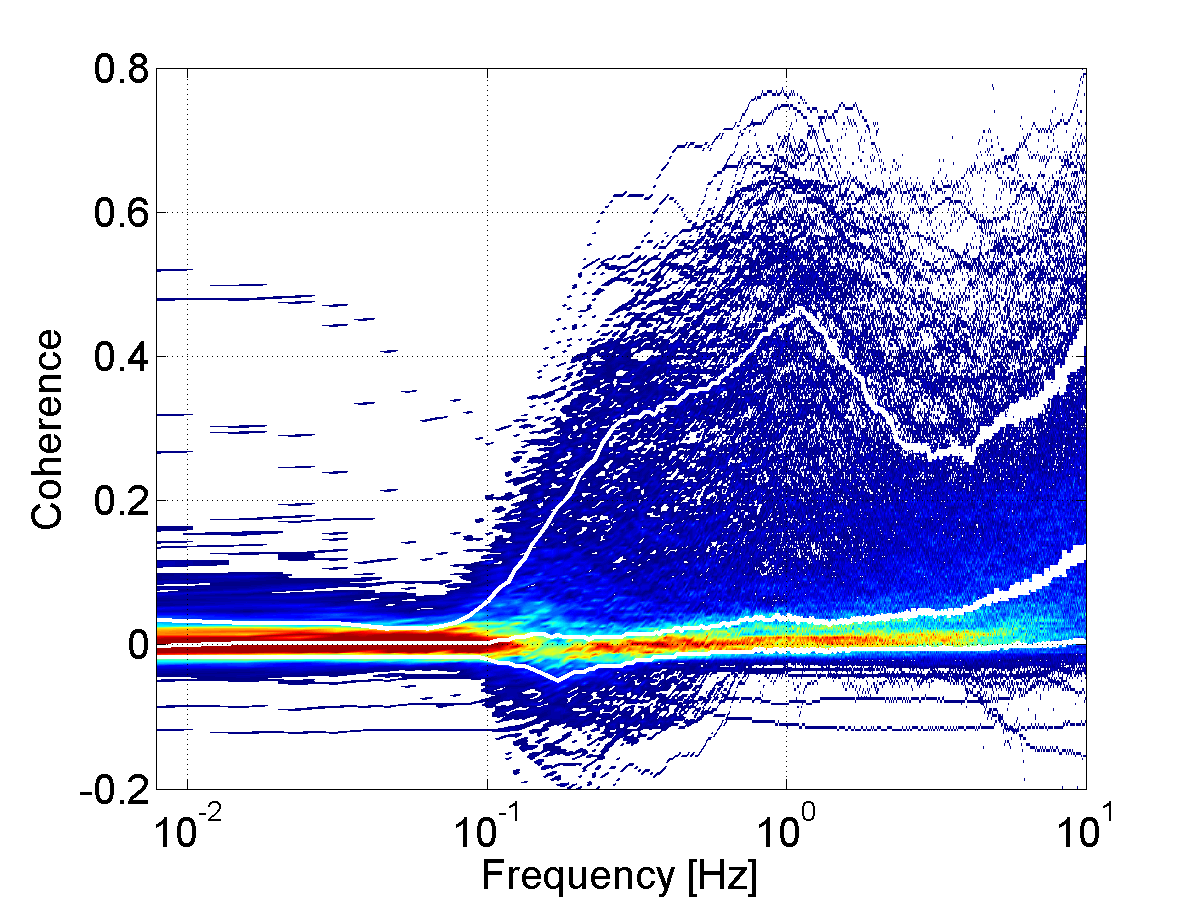}
 \caption{The plots show temporal correlations of seismic spectra with earthquake and wind-speed time series. The earthquake time series were calculated as event-magnitude time series weighted by the inverse epicentral distance to each seismic station. The plot shows that highest correlation is observed just below 0.1\,Hz. The wind-speed correlation is calculated for US stations only. The plot shows that a fraction of 0.1 of all seismic stations exceed a correlation of 0.5 at 1\,Hz.}
 \label{fig:TemporalCorrelation}
\end{figure}

We used the seismic spectra to correlate with other data sets. Spatial correlations were calculated between seismic spectra and topography and population density as shown in figure \ref{fig:SpatialCorrelation}, whereas temporal correlations are evaluated between seismic spectra and wind speed and earthquake data as shown in figure \ref{fig:TemporalCorrelation}. 

We have acquired surface topography data covering the United States from the National Geodetic Survey. The elevation data is provided in blocks of $1\,\deg$ latitude $\times$ $1\,\deg$ longitude with 1 second resolution. The correlation is then evaluated for each frequency between 10th, 50th and 90th percentiles of station spectra and station elevations, resulting in the left plot of figure \ref{fig:SpatialCorrelation}. The coherence above 1\,Hz is as one would expect since high-elevation stations are typically more distant from cities with high population. The strong negative coherence at frequencies of the oceanic microseisms can be explained by the fact that many low-elevation stations in the US were close to the west coast where oceanic microseisms are stronger. This picture may change over the next years when the TA moves into the eastern part of the US that has higher population density and smaller elevation.
It should be noted that since the majority of stations in the US were or are part of the TA moving from the west to the east, US-wide changes of average seismic spectra over time could in principle contribute to the topographic correlation (e.~g.~if seismic spectra happened to be weak everywhere in the US while the TA was mostly located in the Rocky Mountains). However, the US-wide average spectral density as a function of time does not show any significant trends at any frequency so that we can conclude that the result in figure \ref{fig:SpatialCorrelation} is in fact determined by spatial correlations.

We have also acquired population numbers and locations for the 10,000 highest populated cities throughout the world (including cities with less than 50,000 citizens) to evaluate a correlation between seismic spectra and population numbers. We first identified all cities that lie within 50\,km to each seismic station. A station that is more than 50\,km away from any city was excluded. As the intention is to evaluate the correlation individually for various regions in the world, we restrict the analysis to Japan, Europe and the US, which all have sufficiently high numbers of seismic stations and population density. For each seismic station, we divide the population of every city within 50\,km by its distance to the seismic station and add these numbers. These sums are then correlated for all frequencies with the logarithms of the 90th percentiles of seismic spectra. The result can be seen in the right plot of Fig.~\ref{fig:SpatialCorrelation}. We found that whereas the absolute coherence values depend significantly on details about how a figure-of-merit is calculated from city populations, the shape of the three curves only depends weakly on it. This is true for the generic feature that coherence increases with frequency, but also for more special features like the drop of coherence in Japan around 0.6\,Hz and close to 10\,Hz, and the local coherence maxima of Europe and Japan around 0.15\,Hz. The explanation for the decrease of coherence in Japan at higher frequencies is that many F-net stations are located tens of meters underground, which prevents higher-frequency anthropogenic noise to reach the seismometers. We have no conclusive explanation for why seismic spectra in Japan are weaker around 0.6\,Hz when the stations are closer to urban population. One possible cause could be that most F-net stations lie simultaneously in regions with large cities and distant from the coast. In contrast, the high-population samples of most US seismic stations come from the west coast (although the TA is about to enter the high-population regions in the east within the next years). In general, coherence values below 1\,Hz are too small for an unambiguous interpretation of the results.

We calculated a temporal correlation between the occurrence of earthquakes and seismic spectra. We have acquired information about global earthquakes for the years 1998-2011 from the United States Geological Survey's National Earthquake Information Center ``Preliminary Determinations of Epicenters'' project. To construct a time series of global seismic activity, we first exclude all events with magnitudes smaller than 4.4. This value was chosen as the histogram of all earthquake events indicates that a significant fraction of events remains undetected (due to regional gaps in seismic networks) when event magnitude is smaller than 4.4. Therefore, we excluded all these events to avoid a bias from local network density. 
For each seismic station, a time series of earthquake events is generated, splitting each day into three hour segments. More specifically, the magnitudes are first converted from logarithmic to linear scale, then a time series is formed for each seismic station individually by dividing all linear magnitudes by the (great-circle) distance to the epicenters. These values are summed over all earthquakes that occurred within the same 3-hour stretch. This event time series is then correlated with the time series of seismic spectra for each frequency. The coherence spectrum of each station contributes to a histogram that is shown in the left plot of figure \ref{fig:TemporalCorrelation}. The plot also contains the 10th, 50th and 90th coherence percentiles. For example, the 10th percentile is very close to zero coherence, which means that less than 10\% of all seismic stations are weakly affected by earthquakes. Many of these stations lie far from active earthquake zones and are located in regions with strong ambient noise at low-frequency so that many earthquake events leave no trace in 3-hour averaged spectra. Coherence with all percentiles peaks around 80\,mHz. The 90th percentile has a second weaker local maximum around 0.5\,Hz.
\begin{figure}[t]
\centering
 \includegraphics[width=5in]{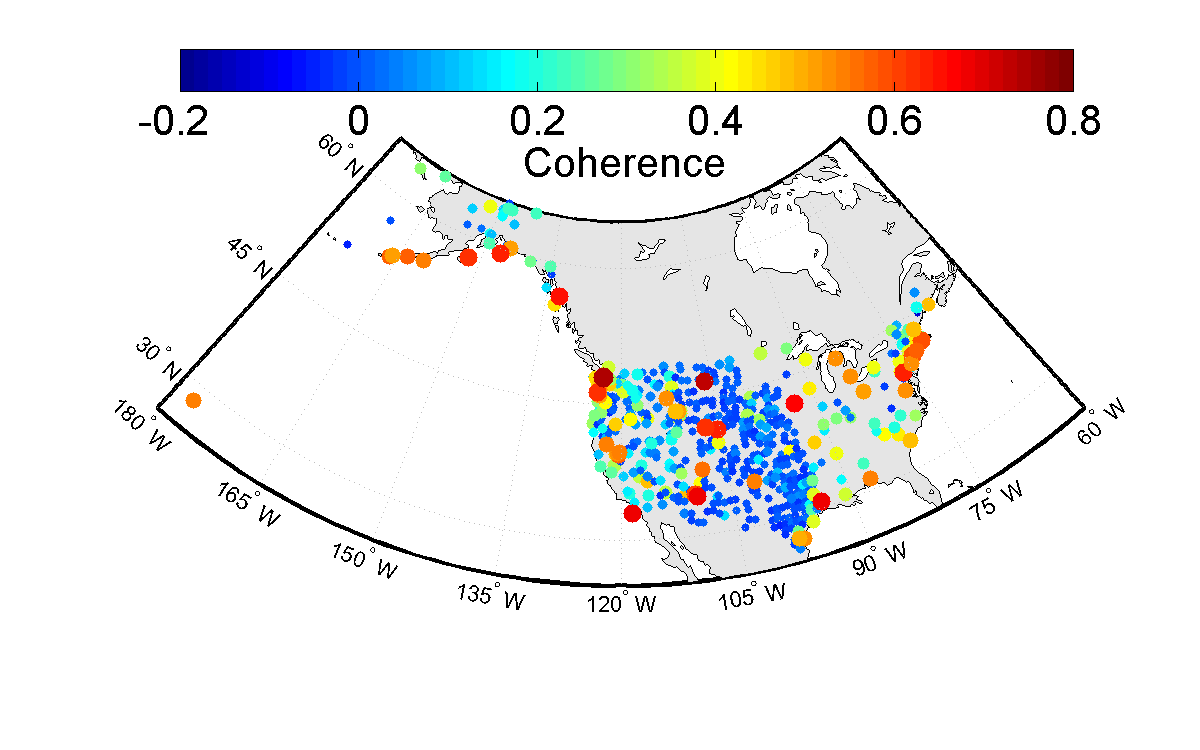}
 \caption{The plot shows the location of the 587 seismic stations that were used for the wind-correlation analysis. The marker sizes and colors both indicate the coherence value between wind speeds and seismic spectra at 1\,Hz. Most stations with coherence larger than 0.5 can be found at the coast. Stations deeper inside the continent typically have very small coherence values, but there are also some high-coherence outliers.}
 \label{fig:windmap}
\end{figure}

As a final example, we present correlations between wind speeds and seismic spectra. The wind data was obtained from the National Climatic Data Center. The correlation is calculated in a similar fashion to the earthquake correlation. Wind time series were generated for each wind station that is collocated (within 50\,km) with a US seismic station. Wind data is first averaged over 3-hour stretches to match the sampling rate of the seismic spectra. The correlation is then evaluated for each frequency bin of the seismic spectra. The spectral coherence at each station contributes to a histogram that is shown in the right plot of figure \ref{fig:TemporalCorrelation}. For all stations, there is no significant coherence below 0.1\,Hz. Although coherence was anticipated, it is certainly surprising that some stations exhibit very strong coherence between seismic noise and wind speeds above 0.1\,Hz. For example, as indicated by the 90th percentile, a fraction 0.1 of all stations exceeds coherence values of about 0.5 at 1\,Hz and 10\,Hz. At some stations wind seems to be the dominant source of seismic noise over a wide range of frequencies. This led us to investigate the location of these stations systematically. Figure \ref{fig:windmap} shows the location of seismic stations that were used for the wind-correlation analysis. The marker sizes and colors both indicate the coherence value at 1\,Hz. Most stations with high wind coherence can be found at the coast. Stations deeper inside the continent typically have very small coherence values, but there are also some high-coherence outliers.

\section{Conclusion}
\label{sec:Conclude}
We presented a seismic analysis based on a new global data set of seismic spectra. Data from more than 3500 seismic broadband seismometers have been processed and stored as 3-hour spectral averages. The processing is complete for the past 5 years; in the future, it will be completed to fill years further back into the past. The paper shows a number of results that can be obtained from the data with relative ease. Stations were grouped geographically to obtain continent specific spectral histograms shown in figure \ref{fig:ContinentPDFs}. The 3-hour time resolution also allowed us to evaluate globally averaged diurnal variations that were presented in figure \ref{fig:diurnal}. 

Comparison to other data sets provides interesting connections and insight to sources of seismic noise. We presented two examples for spatial correlations and two examples for temporal correlations. Spatial correlations were evaluated between seismic spectra and station elevation, as well as between seismic spectra and proximity to urban settlements. Figure \ref{fig:SpatialCorrelation} shows the results that were either subdivided into correlations with different seismic percentiles or into different continents. Results for two temporal correlation measurements were presented in figure \ref{fig:TemporalCorrelation}. The first example is the correlation in time between earthquakes using magnitude weights depending on epicentral distances to all stations. The second example is the correlation between wind speeds and seismic spectra. Especially the histogram of wind correlation shows that some stations exist with surprisingly high coherence values above 0.1\,Hz.

Many more results can easily be obtained from the spectral data. For example, Google Earth files have been generated showing how microseismic peak amplitudes evolve over months in regions with dense seismic networks like Japan, USA or Europe. Ambient seismic noise has been studied in similar ways using Google Earth as interface. Examples can be downloaded at http://www.ligo.caltech.edu/\%7Ejharms/data/GoogleEarth/. In the future, when the data processing is complete for a few decades, the next steps will be to study seismic bands globally over long periods of time and link them with global climatic evolution. Figure \ref{fig:GlobalSeismicAverage} already indicates that average energy in various seismic bands can change significantly within times as short as a year. 

\section{Acknowledgments}

The authors gratefully acknowledge the support of the United States National Science Foundation for the construction and operation of the LIGO Laboratory. M.~C.'s work was funded by the NSF through the California Institute of Technology's Summer Undergraduate Fund.

We thank S.~Anderson for providing the computational resources that made this analysis possible.

Global Seismographic Network (GSN) is a cooperative scientific facility operated jointly by the Incorporated Research Institutions for Seismology (IRIS), the United States Geological Survey (USGS), and the National Science Foundation (NSF). The IRIS DMS is funded through the National Science Foundation and specifically the GEO Directorate through the Instrumentation and Facilities Program of the National Science Foundation under Cooperative Agreement EAR-0552316. We would like to thank Chad Trabant for significant guidance in the download of IRIS data. 

We used the broadband data recorded in the F-net network operated by National Research Institute for Earth Science and Disaster Prevention (NIED). 

We thank the data suppliers stated in the list on http://www.orfeus-eu.org/Data-info/vebsn-contributors.html that contribute to the Virtual European Broadband Seismograph Network (VEBSN). Data access is provided through the Orfeus Data Center.

All plots in this paper were generated with Matlab including the Matlab Mapping Toolbox for the two maps.

%\raggedright
\bibliographystyle{unsrt}
\bibliography{references}

\end{document}